# Use of a Bent Crystal with a Decreasing Curvature to Increase the Efficiency of the Extraction and Collimation of a Beam in an Accelerator

I. A. Yazynin, V. A. Maisheev, and Yu. A. Chesnokov
*Institute for High Energy Physics, Protvino, Moscow region, 142281 Russia*

It has been shown that the use of a bent crystal with a variable curvature radius makes it possible to reduce the fraction of dechanneled particles by an order of magnitude. This effect enables the strong reduction of the particle density at the edge of a collimator or the partition of a septum upon the multiturn extraction of a beam from a ring accelerator. In particular, the beam extraction efficiency at the U-70 synchrotron and Large Hadron Collider can be increased to 95 and 99.65%, respectively.

At present, the collimation of a circulating beam using coherent phenomena in oriented crystals is examined at several large accelerators. Pioneering works [1–3] at the U-70 IHEP accelerator indicate that, in short bent silicon crystals, channeling can increase the efficiency of the extraction and collimation of the beam up to 85%. This possibility has been confirmed at the SPS collider (CERN) [4] and at the Tevatron (Fermilab) [5]. In view of the commissioning of the Large Hadron Collider (LHC) and the problem of an increase in its luminosity, the problem of improving the beam collimation efficiency is of particular importance [6]. Promising experiments on the collimation of the beam using the reflection of particles in crystals were performed [7–10]. Interesting proposals concerning the use of the reflection of particles in crystals with increasing curvature were formulated in [11, 12].

In this work, we propose a method of improving collimation or extraction of the circulating beam using channeling in a short bent crystal with decreasing curvature, which is technically simpler. In this case, only one crystal is necessary, whereas several well oriented crystals are necessary when reflection is used. Channeling in crystals with variable curvature was studied in several experimental works [13, 14]. Low dechanneling was observed in crystal sections with decreasing curvature. This problem for the case of an extended crystal was analyzed theoretically in [15], where it was noted that crystals with decreasing curvature is applicable for decreasing radiation loads on the elements of an accelerator due to the suppression of the dechanneling process.

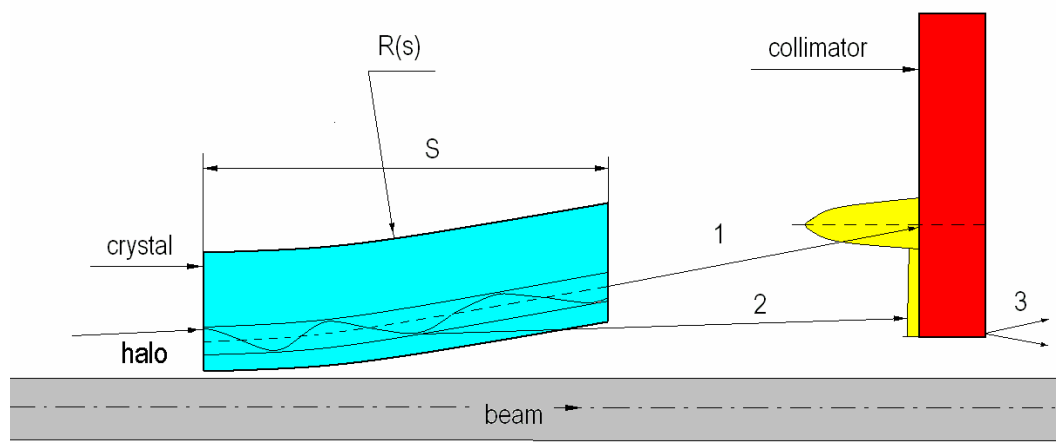

Fig. 1. Layout of beam collimation using the channeling crystal.

Figure 1 shows the layout of the use of a short bent crystal for beam collimation in an accelerator. Particles of a circulating beam increase the amplitude of transverse oscillations due to numerous effects, such as scattering on a residual gas, the effect of nonlinearities, and the scattering at the interactions point. As a result, the beam halo where particles fall on the leading edge of the crystal appears. Due to the channeling effect, the majority of the beam halo (fraction *1*) is deeply deflected to the collimator (absorber). Only several percents of particles are deflected at an incomplete angle due to dechanneling (fraction *2*), which leads to radiation losses on the accelerator (scattering and secondary particles *3*). A crystal with decreasing curvature can considerably reduce the dechanneling fraction of particles getting on collimator edge and to lower the losses.

The essence of the suppression of dechanneling can be illustrated by representing the motion of particles on the coordinate–angle phase plane (details of this approach can be found in [15]). We consider a crystal with a continuously varying curvature radius with a length of s along the beam. The dependence of the radius on the length is specified as $R(s) = R_1(1 + sC_r/S)$, where $C_r$ is the curvature variation coefficient. At the input of the crystal, the radius is $R(0) = R_1$ and at the output of the crystal with a length of S, it increases to the maximum value $R(S) = R_1(1 + C_r)$. The bending angle and average curvature radius of the crystal were defined as

$$\alpha = \int_0^S \frac{ds}{R(S)} = \frac{S}{R_1} \cdot \frac{\ln(1+C_r)}{C_r}, \quad R_0 = \frac{S}{\alpha} = \frac{R_1}{\chi(C_r)}, \quad \text{where} \quad \chi(C_r) = \frac{\ln(1+C_r)}{C_r}.$$

As an example, we consider the motion of protons on the phase plane in a (110) silicon crystal (the same crystals will be used below). For 7 TeV protons, which correspond to the LHC, the beam is collimated by a crystal 3 mm in length with an average radius of 100 m and the coefficient $C_r = 1$. The optimum length of the crystal and its average bending radius were calculated in [16]. Closed line *1* in Fig. 2 is the boundary of the phase region of trajectories of particles that are available for channeling at the beginning of the crystal (the input acceptance). Particles in this region undergo periodic oscillations around the equilibrium orbit $x_{or}$.

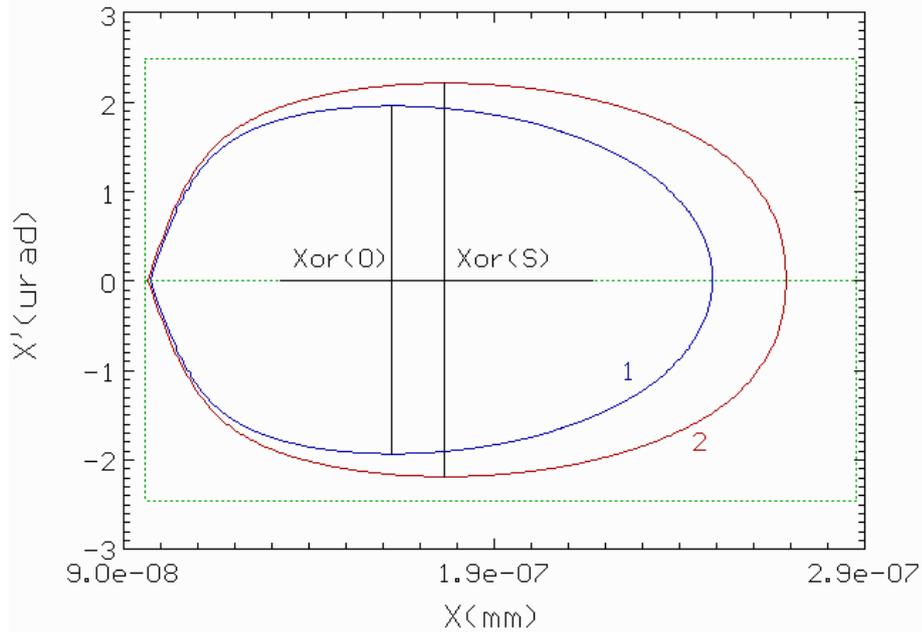

Fig. 2. Beam channeling region in the bent silicon crystal with decreasing curvature at the (*1*) beginning and (*2*) end of the crystal.

The particles moving in the crystal are scattered from the electrons and nuclei of the lattice and, hence, some of them leave the channeling mode (dechanneling process). At the same time, due to an increase in the curvature radius along the particle trajectories, the available channeling region expands (line *2*), particles move away from the maximum scattering boundary, and the fraction of dechanneled particles decreases. For the strict quantitative description of the interaction of protons with a bent crystal with variable curvature, we used SCRAPER software [17], which is based on the Monte Carlo method. This program was tested by comparing its results with the experimental data [18] on the channeling of 400 GeV proton beam in a short silicon crystal. The calculated fraction of dechanneled particles is in agreement with the value measured in [18] within 1%.

Then, the single passage of protons through the crystal with a decreasing curvature was simulated for two energies of particles in the beam. The first calculation was performed for 70 GeV protons, which are available at the U-70 accelerator (IHEP). The length of the crystal optimal for the most efficient multiturn extraction and collimation in this case is $S \approx 0.4$ mm at the deflection angle $\alpha = 400$ μrad [16]. Figure 3a shows the distribution of the narrow beam deflected by the crystal for various crystal curvature variation coefficients from uniformly bent ($C_r = 0$) to the maximum value $C_r = 4$.

The channeled fraction (right peak) decreases from 77 to 63% with an increase in the curvature gradient, whereas the fraction of protons that are not initially involved in the channeling mode (left peak) increases from 20 to 35% due to a decrease in the input acceptance. The density of the dechanneled protons (the region between the peaks) decreases by almost two orders of magnitude at the deflection angle $x' = 0.2$–$0.25$ mrad, which corresponds to particles that fall onto the septum or collimator edge.

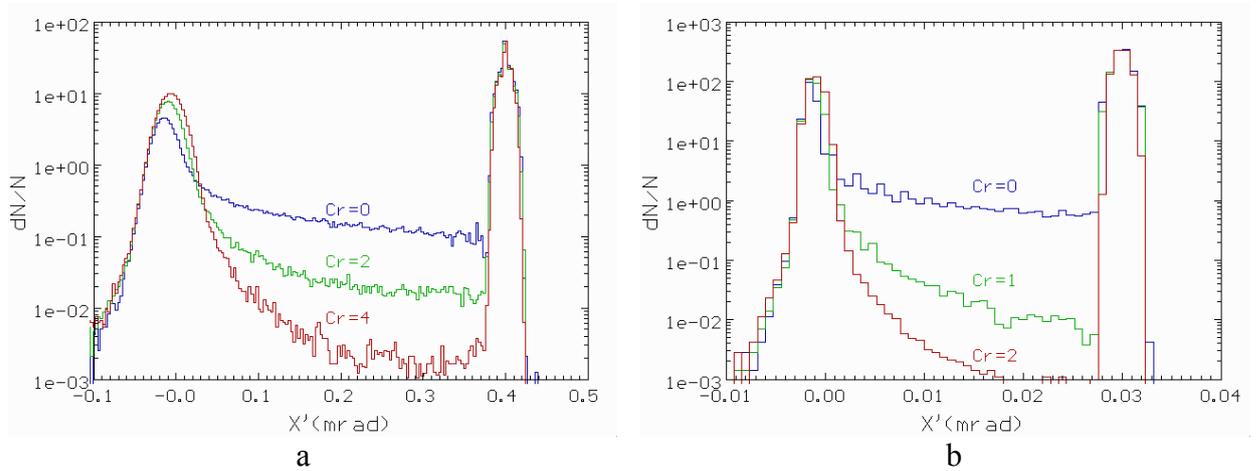

a          b

Fig. 3. Distribution of protons with the energy $E =$ (a) 70 GeV and (b) 7 TeV in the angular deviation at the output of the Si(110) crystal.

Similar angular distributions of 7 TeV protons at the output of the crystal with the parameters $\alpha = 30$ μrad and $L_c = 3$ mm optimal for collimating the beam at the LHC are shown in Fig. 3b. The density of the dechanneled fraction decreases by two orders of magnitude at the edge of the main collimator $x' = 15$–$20$ μrad. This makes it possible to strongly reduce the number of protons that leave the collimators and are lost in the ring. The fraction of channeled protons (right peak) decreases from 83 to 74% with an increase in the curvature gradient. For this reason, to optimize the collimation scheme or the beam extraction scheme at the LHC, simulation should take into account both the multiturn passage of particles through the crystal in the accelerator and losses of particles on the collimator and crystal.

It was shown previously [16] that the crystal with a constant curvature radius can throw about 99% of particles fairly deep into the collimator at the LHC. According to the calculations with the SCRAPER program, where the multiturn motion of particles in the accelerator was taken into account, the use of the crystal with a decreasing curvature allows an additional increase in the efficiency of the collimation system. Figure 4 shows the calculated distributions in the coordinate at the main collimator of the LHC for the crystal with variable curvature.

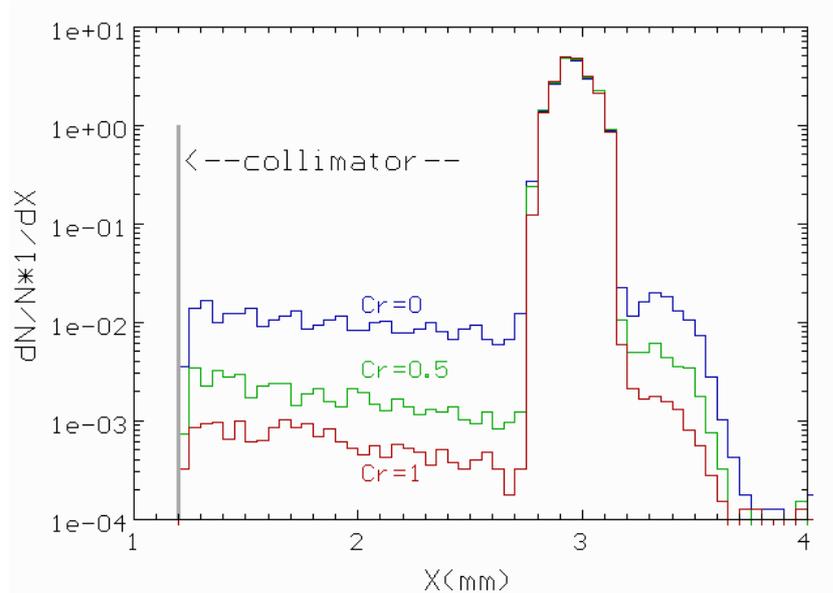

Fig. 4. Distribution of particles at the scraper of the LHC for various curvatures Cr.

The fraction of channeled protons penetrates deep into the collimator at a distance of 1.5 mm from its edge and is completely absorbed. Only the fraction of dechanneled protons achieves the edge of the collimator. It can be seen in the distributions that the density of particles at the edge of the collimator decreases by more than an order of magnitude for the crystal with the decreasing curvature. The minimum of the total losses on the crystal and collimator is observed at the curvature coefficient $C_r = 1$. In this case, losses are reduced by a factor of 4 (from 1.3 to 0.35%); i.e., the efficiency of the extraction at a septum of 1 mm reaches 99.65%, which cannot be achieved even for a slow resonance extraction.

The calculations of the multiturn extraction at an energy of 70 GeV for the U-70 accelerator indicate that the crystal with the decreasing curvature can increase the extraction efficiency from 85 to 95%. To conclude, we note that the production of a crystal with a variable bending radius is easy and can be achieved using devices applied in [3].

This work was supported by the Russian Foundation for Basic Research, project nos. 08-02-13533-ofi-ts, 08-02-01453-a, and 11-02-90415-Ukr-f-a.